\documentclass[aps,pra,twocolumn,groupedaddress,showpacs,floatfix,superscriptaddress]{revtex4-2}

\usepackage[T1]{fontenc}
\usepackage[utf8]{inputenc}
\usepackage{natbib}
\usepackage{babel}
\usepackage{amsmath}
\usepackage{amssymb}
\usepackage{graphicx}
\usepackage[unicode=true, bookmarks=false, breaklinks=false,pdfborder={0 0 1},backref=false,colorlinks=false]
 {hyperref}

\usepackage{breakurl}

\makeatletter

 \@ifundefined{textcolor}{}
 {
   \definecolor{BLACK}{gray}{0}
   \definecolor{WHITE}{gray}{1}
   \definecolor{RED}{rgb}{1,0,0}
   \definecolor{GREEN}{rgb}{0,1,0}
   \definecolor{BLUE}{rgb}{0,0,1}
   \definecolor{CYAN}{cmyk}{1,0,0,0}
   \definecolor{MAGENTA}{cmyk}{0,1,0,0}
   \definecolor{YELLOW}{cmyk}{0,0,1,0}
 }
\usepackage{bbm, dsfont}
\usepackage{bm}
\usepackage{amsfonts}
\hypersetup{colorlinks=true, citecolor=blue,linkcolor=blue, filecolor=blue, urlcolor=blue}

\newcommand{\ket}[1]{\left| #1 \right>} 
\newcommand{\bra}[1]{\left< #1 \right|}

\makeatother

\begin{document}

\title{Quantum coherence induced by magnons}

%%%%%%%%%%%%%%%%%%%%%%%%%%%%%%%%%%%%%%%%%%%%%%%
%%%%%%%%%%%%%%---Authors---%%%%%%%%%%%%%%%%%%%%
%%%%%%%%%%%%%%%%%%%%%%%%%%%%%%%%%%%%%%%%%%%%%%%
\author{V. C. Vieira}
\affiliation{Centro de Ci\^ encias Naturais e Humanas, Universidade Federal do ABC, Avenida dos Estados 5001, 09210-580, Santo Andr\'e, S\~ao Paulo, Brazil}

\author{F. M. de Paula}
\affiliation{Centro de Ci\^ encias Naturais e Humanas, Universidade Federal do ABC, Avenida dos Estados 5001, 09210-580, Santo Andr\'e, S\~ao Paulo, Brazil}

\date{\today}
%%%%%%%%%%%%%%%%%%%%%%%%%%%%%%%%%%%%%%%%%%%%%%%
%%%%%%%%%%%%%%---Abstract---%%%%%%%%%%%%%%%%%%%
%%%%%%%%%%%%%%%%%%%%%%%%%%%%%%%%%%%%%%%%%%%%%%%
\begin{abstract}
We analyze quantum coherence generated by non-interacting magnons in a ferromagnetic spin chain described by the isotropic Heisenberg model. The exact expression derived for the reduced density operator of an arbitrary subsystem reveals that single-site coherence vanishes, while maximal coherence arises in a pure single-mode magnon state. Remarkably, this maximum depends solely on the number of coherently superposed microstates. Based on this property, we introduce coherence-based definitions of entropy, temperature, and heat capacity, establishing a formal analogy with standard thermodynamic quantities.

\end{abstract}

\maketitle
%%%%%%%%%%%%%%%%%%%%%%%%%%%%%%%%%%%%%%%%%%%%%%%
%%%%%%%%%%%%---Introduction---%%%%%%%%%%%%%%%%%
%%%%%%%%%%%%%%%%%%%%%%%%%%%%%%%%%%%%%%%%%%%%%%%
\section{Introduction}
Quantum coherence (QC) expresses the quantum principle of state superposition inherent in quantum theory, a property essential for phenomena such as interference~\cite{Feynman65}, tunneling~\cite{Merzbacher02}, and particle entanglement~\cite{Amico08,Hordocki09}. Its non-classical nature generated skepticism about its validity in the early days of quantum theory, as illustrated, for example, by Schor\"ondiger's cat paradox~\cite{Schrodinger35}. However, experiments have shown that QC is not only a physical reality at subatomic and atomic scales, but can also manifest in mesoscopic systems~\cite{Leggett02,Caldeira14}. Notable examples include the tunneling of supercurrents in a SQUID~\cite{Friedman00,Mooij10},  tunneling of magnetization in magnetic materials~\cite{Chudnovsky88, Stamp92, Friedman10, Awschalom92, Luis00}, quantum superposition of vibrational states in microscopic mechanical resonators~\cite{OConnell10,Cho10}, interference with organic macromolecules~\cite{Arndt99, Gerlich11, Fein19}, interference between Bose-Einstein condensates (BECs)~\cite{Andrews96}, among many others.

The purpose of research on QC goes beyond probing the boundaries of Quantum Theory. From a practical perspective, QC is an invaluable physical resource, essential for performing tasks in quantum computing and quantum information~\cite{Hillery16, Ma16, Matera16, Shin17, Giorda18, Bu2017}, with potential applications in a wide range of other promising fields, such as quantum thermodynamics~\cite{Goold16, Landi21} and quantum biology~\cite{Cao2020}. Despite its fundamental importance, the development of a rigorous theory of QC as a physical resource only began in 2014~\cite{Baumgratz14}. This advancement allowed a precise value to be assigned to the QC encoded in a quantum state and spurred research aimed at elucidating more quantitative aspects of QC exhibited by various physical systems. In particular, intensive studies on measures of QC in condensed matter systems have been conducted, revealing connections between QC and quantum properties of this class of many-body systems~\cite{Streltsov17, Hu18}. For example, it has been shown that the off-diagonal long-range order (ODLRO) exhibited by a superconducting state (a condensate of electron pairs) described by the Hubbard model is directly related to the QC of this state, as measured by the so-called 
$l_1$-norm of coherence (a geometric quantifier of QC)~\cite{Hu18}. Similar connections are expected in other condensed matter systems, such as superfluids, ferromagnets, and antiferromagnets.

In magnetically ordered systems, information can be encoded and transported through magnons~\cite{Chumak15, Andrianov14}, that is, the quanta of collective magnetic moment excitations (spin waves)~\cite{Kittel63}. The emerging field that studies the generation and manipulation of magnons is known as magnonics (a subfield of spintronics)~\cite{Yuan22}. One of the main interests in magnons lies in their electrical neutrality, which helps to reduce dissipation and decoherence effects~\cite{Kajiwara10, Bunkov10}. Furthermore, BECs of magnons are now being experimentally observed at room temperature~\cite{Schneider20, Divinskiy21}. This type of BEC can become an attractive source of QC and, consequently, play a central role in the development of new quantum technologies, as well as in tests of the validity of quantum theory at the macroscopic level.

In this work, we investigate how QC emerges in non-interacting magnon states of a ferromagnetic Heisenberg spin chain. By deriving the reduced density matrix of an arbitrary subsystem, we show that QC is maximized when all magnons occupy the same momentum state. Remarkably, the maximum coherence is governed by the number of orthogonal basis states coherently superposed. Building on this observation, we propose a computationally efficient coherence-based quantity analogous to the Boltzmann entropy, as well as effective definitions of temperature and heat capacity. As an application, we investigate the coherence thermodynamics of a single-mode magnon state, highlighting the correspondence with the standard thermodynamics of a two-level Boltzmann gas.

This paper is organized as follows. In Section~\ref{II}, we introduce the isotropic Heisenberg model for a ferromagnetic spin chain and derive the reduced density operator of an arbitrary subsystem in the dilute magnon regime. In Section~\ref{III}, we discuss the theory of QC, identify the conditions that optimize it for magnon states, and additionally propose a coherence-based definition of entropy. In Section~\ref{IV}, we develop a thermodynamic framework for QC and illustrate it through an application. Finally, Section~\ref{V} presents our conclusions and possible directions for future research.

%%%%%%%%%%%%%%%%%%%%%%%%%%%%%%%%%%%%%%%%%
%%%%%%%%%%%%---Magnon states--%%%%%%%%%%%
%%%%%%%%%%%%%%%%%%%%%%%%%%%%%%%%%%%%%%%%%
\section{Magnon states}
\label{II}
Let us consider a one-dimensional chain of $N$ spin-1/2 particles, described by the isotropic ferromagnetic Heisenberg model with nearest-neighbor interactions, whose Hamiltonian is given by~\cite{Morimae05}:
\begin{equation}
H = -J \sum_{l=1}^{N} \sigma_l \cdot \sigma_{l+1}
\end{equation}
where $J>0$ denotes the exchange interaction and the symbol $\sigma_l = (\sigma_{l}^x, \sigma_{l}^y, \sigma_{l}^z)$ represents the spin operator at site $l$. One of the ground states of $H$ is given by $\ket{\downarrow^{\otimes N}}$, which physically represents a state with total magnetization pointing downward. Starting from this ordered state, one can induce a defect (a spin-up) propagating along the chain as a wave. This spin excitation, which represents a magnon in the system, is described by the following quantum state~\cite{Morimae05}:
\begin{equation}
\ket{\psi_{Nk}} = M_k^\dagger\ket{\downarrow^{\otimes N}}=\frac{1}{\sqrt{N}} \sum_{l=1}^{N} e^{ikl} \ket{l}
\end{equation}
where
\begin{equation}
M_k^\dagger = \frac{1}{\sqrt{N}} \sum_{l=1}^{N} e^{ikl} \sigma_{l}^+
\end{equation}
defines the magnon creation operator with $\sigma_{l}^+ =(\sigma_{l}^x + i\sigma_{l}^y)/2$ representing the raising operator acting on site $l$: $\sigma_l^+\ket{\downarrow...\downarrow^{(l)}...\downarrow}=\ket{\downarrow...\uparrow^{(l)}...\downarrow}\equiv\ket{l} $. This is an eigenstate of $H$ with energy 
\begin{equation}\label{emag}
 \epsilon_k = 8J\sin^2(k/2)   
\end{equation}
where $k$ denotes the magnon momentum (wave number). In particular, $k$ takes $N$ different values for periodic boundary conditions: $k=2\pi n/N$ with $n=0, 1, ..., N-1$ when $\sigma_{N+1}=\sigma_1$. A state with $m$ magnons ($m=1,2,...,N$) is obtained by applying the magnon creation operator $m$ times to the ground state:
\begin{equation}\label{ket}
\ket{\psi_{Nm\textbf{k}}}=G_{Nm\textbf{k}} \sum_{\textbf{l}\in C_m^N} f_{\textbf{kl}}\ket{\textbf{l}}
\end{equation}
where
\begin{equation}
G_{Nm\textbf{k}}= \left(\sum_{\textbf{l}\in C_m^N} |f_{\textbf{kl}}|^2\right)^{-1/2}
\end{equation}
is the normalization factor, with
\begin{equation}
f_{\textbf{kl}}=\sum_{\pi\in S_m} e^{i\textbf{k}_\pi\cdot\textbf{l} }.
\end{equation}
In these expressions, $\textbf{l}=\{l_1,...,l_m\}$ is a list that indicates the sites with spin-up, which belongs to the set $C^N_m$, i.e., the set of $m$-combinations of $N$ distinct elements. The vector $\textbf{k}=\{k_1,...,k_m\}$ provides the wave number distribution and $\textbf{k}_\pi=\{k_{\pi(1)},...,k_{\pi(m)}\}$ a vector formed by a permutation of the elements of $\textbf{k}$, with $\pi$ belonging to the symmetric group $S_m$, i.e., the set of all permutations of $m$-distinct elements. The cardinalities (number of elemens) of the sets $C^N_m$ and $S_m$ are, respectively,
\begin{equation}
    \left| C^N_m\right| =\binom{N}{m}=\frac{N!}{m!(N-m)!}
\end{equation}
and
\begin{equation}
    \left|S_m\right|=m!.
\end{equation}

For low magnon densities ($m \ll N$), Eq.~\eqref{ket} represents an approximate eigenstate of $H$, with magnons acting as bosons. In this case, the density operator in the orthonormal basis $\{\ket{\mathbf{l}}\}$ is given by
\begin{equation}
\rho_{Nm\textbf{k}}=\ket{\psi_{Nm\textbf{k}}}\bra{\psi_{Nm\textbf{k}}}= \sum_{\textbf{l},\textbf{l}'\in C_m^N}\rho_{Nm\textbf{k}}^{\textbf{l\,l}'}\ket{\textbf{l}}\bra{\textbf{l}'}
\end{equation}
with
\begin{equation}
\rho_{Nm\textbf{k}}^{\textbf{l\,l}'}=|G_{Nm\textbf{k}}|^2 f_{\textbf{kl}}f_{\textbf{kl}'}^*
\end{equation}
representing the matrix elements of the  pure state $\rho_{Nmk}$. Now, let us partition the chain into two arbitrary subsystems, $A$ and $B$, consisting of $n$ and $N-n$ sites, respectively. The density operator associated with the subsystem $A$ can be written as
\begin{equation}\label{rho}
\rho_{nNm\textbf{k}}=\text{tr}_B[\rho_{Nm\textbf{k}}]=\sum_{q\in Q}\,\sum_{\textbf{l}_A, \textbf{l}_A'\in C_q^n} \rho_{nNm\textbf{k}}^{\textbf{l}_A\textbf{l}_A'}\ket{\textbf{l}_A}\bra{\textbf{l}_A'}
\end{equation}
where
\begin{equation}\label{elements}
\rho_{nNm\textbf{k}}^{\textbf{l}_A\textbf{l}'_A}=\sum_{\textbf{l}_B\in C_{m-q}^{N-n}} \rho_{Nm\textbf{k}}^{\textbf{l}_A\cup \textbf{l}_B\,\textbf{l}_A'\cup \textbf{l}_B} 
\end{equation}
are the corresponding matrix elements. Here, $\mathbf{l}_{A(B)}$ is a vector whose entries specify the sites within subsystem $A (B)$ that exhibit spin-up and
\begin{equation}
Q=\{q\in\mathbb{Z}\mid\max[0,m-(N-n)]\leq q\leq \min[n,m]\}  
\end{equation}
denotes the set of all admissible values $q$ of magnons in subsystem $A$. In the special case of a single-mode magnon state, where all wave numbers take on the same value $k$, i.e., $\mathbf{k}=\mathbf{k}_0=\{k,...,k\}$, the matrix elements described in Eq.(\ref{elements}) assume the form
\begin{equation}
\rho_{nNm\textbf{k}_0}^{\textbf{l}_A\textbf{l}'_A}=\frac{\left| C^{N-n}_{m-q}\right|}{\left| C^{N}_{m}\right|} e^{i\textbf{k}_{A0}\cdot(\textbf{l}_A-\textbf{l}_A')}
\end{equation}
where $\mathbf{k}_{A0}$ is a subset of $\mathbf{k}_0$ containing $n$ elements. Consequently, the density operator of subsystem $A$ reduces to an average of a pure single-mode magnon state over the magnon number:
\begin{equation}\label{maxred}
\rho_{nNm\textbf{k}_0}=\sum_{q\in Q} p(q)\ket{\psi_{nq\textbf{k}_0}}\bra{\psi_{nq\textbf{k}_0}}
\end{equation}
where
\begin{equation}
p(q)=\frac{\left| C^{N-n}_{m-q}\right|\left| C^{n}_{q}\right|}{\left| C^{N}_{m}\right|}  
\end{equation}
is the hypergeometric distribution~\cite{Feller1968}, representing the probability of finding $q$ magnons within subsystem $A$, and
\begin{equation}\label{bec}
\ket{\psi_{nq\textbf{k}_0}}=\frac{1}{\sqrt{|C^n_q|}}\sum_{\textbf{l}_A\in C_q^n} e^{i\textbf{k}_{A0}\cdot \textbf{l}_A}\ket{\textbf{l}_A}
\end{equation}
denotes a pure state with $q$ magnons occupying the same mode $k$ in a chain composed of $n$ sites.
%%%%%%%%%%%%%%%%%%%%%%%%%%%%%%%%%%%%%%%%%%%%%%%%%%%%%
%%%%%%%%%%%%%---Quantum coherence--%%%%%%%%%%%%%%%%%%
%%%%%%%%%%%%%%%%%%%%%%%%%%%%%%%%%%%%%%%%%%%%%%%%%%%%%
\section{Quantum coherence}
\label{III}
A measure of quantum coherence (QC) quantifies the degree of superposition present in a quantum state relative to a fixed reference basis. According to the framework established by Baumgratz, Cramer, and Plenio~\cite{Baumgratz14}, a bona fide measure of QC must satisfy four criteria:
\begin{enumerate}
    \item \textit{Non-negativity}: 
  $ C(\rho)\geq 0$ for all $\rho$ with  $C(\delta)=0$ iff $\delta\in\mathcal{I}$, i.e., criterion 1 requires that a valid QC measure should be a non-negative quantity that vanishes for incoherent states, i.e., diagonal states in the reference basis, with $\mathcal{I}$ denoting the set of all incoherent states.
  
    \item \textit{Contractivity under incoherent operations}: $C(\rho)\geq C(\Phi_{\text{I}}[\rho])$, i.e., criterion 2 imposes contractivity under a completely positive trace-preserving map satisfying $\Phi_{\text{I}}[\mathcal{I}]\subset\mathcal{I}$.

     \item \textit{Non-increasing under ISM on average}: $C(\rho)\geq \sum_i p_i C(\rho_i)$ with $p_i=\text{tr}[K_i\rho K_i^{\dagger}]$ and $\rho_i=K_n\rho K_i^{\dagger}/p_i$, i.e., criterion 3 establishes monotonicity under an incoherent selective measurement (ISM) on average described by Kraus operators $\{K_i\}$ such that $K_i\mathcal{I} K_i^{\dagger}\subset\mathcal{I}$.  
     
     \item \textit{Convexity}: $C\left(\sum_i p_i\rho_i\right)\leq\sum_i p_i C(\rho_i)$, i.e., criterion 4 requires a convex function to avoid an increase by mixing states, where $\{\rho_i\}$ denotes any set of states and $p_i$ any probability distribution.
\end{enumerate}
Given a fixed orthonormal basis $\{\ket{j}\}_{j=1}^{d}$ of a $d$-dimensional Hilbert space, an incoherent quantum state takes the form
\begin{equation}\label{incoh}
    \delta=\sum_{j=1}^{d}p_j\ket{j}\bra{j}
\end{equation}
where $p_j$ is a probability distribution. Conversely, a maximally coherent state is a pure state $\ket{\psi_d}$ from which any $d\times d$ density matrix can be deterministically obtained through incoherent operations \cite{Baumgratz14, Peng16}:
\begin{equation}\label{ketmax}
    \ket{\psi_d}=\frac{1}{\sqrt{d}}\sum_{j=1}^{d}e^{i \phi_j}\ket{j},\,\,\,\phi_j\in\mathbb{R}.
\end{equation}
A general geometric measure of QC fulfilling the fundamental criteria can be expressed as the minimal distance $\mathcal{D}$ between $\rho$ and the set of all incoherent states $\mathcal{I}$:
\begin{equation}\label{cohD} C(\rho)=\min_{\delta\in\mathcal{I}}\mathcal{D}(\rho,\delta).
\end{equation}  
Common examples include the relative entropy $S(\rho\|\delta)=\text{tr}[\rho(\ln\delta-\ln\rho)]$ and the $l_1$-distance $\|\rho-\delta\|_{l_1}=\sum_{i,j}|(\rho-\delta)_{ij}|$, where $(\rho-\delta)_{ij}$ are the elements of the matrix of $\rho-\delta$ in the fixed basis. For these cases,  Eq.~\eqref{cohD} leads to the relative entropy of coherence and $l_1$-norm of coherence, respectively:
\begin{equation}\label{cohr}
    C_r(\rho)=S(\rho_I)-S(\rho)
\end{equation}
and
\begin{equation}
    C_{l_1}(\rho)=\|\rho-\rho_I\|_{l_1}
\end{equation}
with
\begin{equation}
    \rho_I=\sum_{j=1}^{d}\rho_{jj}\ket{j}\bra{j}
\end{equation}
representing the incoherent component of $\rho$ (it minimizes Eq.~\eqref{cohD} for both distances). In Eq.~\eqref{cohr}, $S(\rho)=-\text{tr}[\rho\ln\rho]$ denotes the von Neumann entropy and $S(\rho_I)$ its incoherent part (the Shannon entropy). Since von Neumann entropy requires diagonalization for mixed states, the $l_1$-norm of coherence is comparatively easier to compute than the relative entropy of coherence. In the particular case of a maximally coherent state, these QC quantifiers solely depend on the Hilbert space dimension:
\begin{equation}
    C_r(\ket{\psi_d})=\ln d,
\end{equation}
\begin{equation}
    C_{l_1}(\ket{\psi_d})=d-1.
\end{equation}

Let us now analyze the quantum coherence properties of magnon states. Given that the partial trace is an incoherent operation, the general expression in Eq.~\eqref{crho} must reach its minimum for $n=1$. In fact, the density matrix for $n=1$ is diagonal, indicating absence of single-site coherence regardless of the QC quantifier:
\begin{equation}
    C(\rho_{1Nm\textbf{k}})=0
\end{equation}
where
\begin{equation} \rho_{1Nm\textbf{k}}=\rho_{1Nm\textbf{k}}^{00}\ket{0}\bra{0}+\rho_{1Nm\textbf{k}}^{11}\ket{1}\bra{1}.
\end{equation}
On the other hand, a maximally coherent state is attained for a pure single-mode magnon state. Indeed, substituting $\textbf{k}=\textbf{k}_0$ into the pure state given by Eq.~\eqref{ket} yields a state of the form described in Eq.~\eqref{ketmax}:
\begin{equation}\label{bec}
\ket{\psi_{Nm\textbf{k}_0}}=\frac{1}{\sqrt{|C^N_m|}}\sum_{\textbf{l}\in C_m^N} e^{i\textbf{k}_0\cdot \textbf{l}}\ket{\textbf{l}}.
\end{equation}
Consequently,
\begin{equation}
    C_r(\ket{\psi_{Nm\textbf{k}_0}})=\ln|C^N_m|,
\end{equation}
\begin{equation}
    C_{l_1}(\ket{\psi_{Nm\textbf{k}_0}})=|C^N_m|-1.
\end{equation}
It is worth mentioning that these expressions become averages for the corresponding reduced state described in Eq.~\eqref{maxred}:
\begin{equation}
C(\rho_{nNm\textbf{k}_0})=\sum_{q\in Q} p(q)C(\ket{\psi_{nq\textbf{k}_0}}).
\end{equation}
We also derive the expression for the $l_1$-norm of coherence of the general reduced density operator given by Eq.~\eqref{rho}:
\begin{equation}\label{crho}
C_{l_1}(\rho_{nNm\textbf{k}})=\sum_{q\in Q}\sum_{\textbf{l}_A, \textbf{l}_A'\in C^n_q} |\rho_{nNm\textbf{k}}^{\textbf{l}_A\textbf{l}'_A}|-1.
\end{equation}
Furthermore, we propose an additive and computationally efficient coherence measure based on $C_{l_1}$, defined as
\begin{equation}
    C_{\ln}(\rho)=\ln\|\rho\|_{l_1}.
\end{equation}
The $l_1$-norm of $\rho$ can be interpreted as an effective dimension of the quantum state. Since $\|\rho\|_{l_1}=1+C_{l_1}$, this alternative measure is monotonically related to $C_{l_1}$, it vanishes for incoherent states, and coincides with $C_r$ for maximally coherent states. In the case of the arbitrary magnon state, the effective dimension is simply given by
\begin{equation}\label{crho}
\|\\\rho_{nNm\textbf{k}}\|_{l_1}=\sum_{q\in Q}\sum_{\textbf{l}_A, \textbf{l}_A'\in C^n_q} |\rho_{nNm\textbf{k}}^{\textbf{l}_A\textbf{l}'_A}|.
\end{equation}

%%%%%%%%%%%%%%%%%%%%%%%%%%%%%%%%%%%%%%%%%%%%%
%%%%---Coherence thermodynamics---%%%%%%%%%%%
%%%%%%%%%%%%%%%%%%%%%%%%%%%%%%%%%%%%%%%%%%%%%
\section{Coherence thermodynamics}
\label{IV}
In a $2^N-$ dimensional Hilbert space, the maximally coherent state in Eq.~\eqref{bec} belongs to a $|C_m^N| -$ dimensional degenerate subspace corresponding to a macrostate of fixed energy. Thus, in analogy to the standard microcanonical ensemble, the pure single-mode magnon state represents a coherent microcanonical ensemble, where all $|C_m^N|$ coherently superposed microstates possess fixed energy and are equiprobable. In this scenario, quantum coherence plays a role analogous to Boltzmann entropy, characterizing the degree of uncertainty among the accessible states. This perspective naturally suggests the development of a thermodynamic theory of coherence. To introduce coherence-based thermodynamic quantities, let us assume quantum coherence as a state function 
\begin{equation}
C=C(U,\{\lambda_i\})
\end{equation}
with 
\begin{equation}\label{intenerg}
    U=\text{tr}[\rho H]
\end{equation}
representing the internal energy (i.e., the average energy) and $\{\lambda_i\}$  complementary parameters characterizing the quantum state. In analogy to standard thermodynamics, we introduce a coherence temperature and the corresponding coherence heat capacity as
\begin{equation}
\mathcal{T}_C=\left(\frac{\partial C}{\partial U}\right)_{\{\lambda_i\}}^{-1},
\end{equation}
and
\begin{equation}
\mathcal{C}=\left(\frac{\partial U}{\partial \mathcal{T}_C}\right)_{\{\lambda_i\}}.
\end{equation}
The set $\{\lambda_i\}$ is held fixed to isolate the dependence of the coherence on energy, as well as to ensure that the coherence heat capacity is expressed as a function of coherence temperature alone. It is important to mention that some quantum versions of temperature have been defined in terms of von Neumann entropy \cite{Vallejo20,Vallejo21,Alipour21,Alipour22,Choquehuanca2025}, with the set $\{\lambda_i\}$ depending on the quantum definition of heat and work. Remarkably, the inverse of coherence temperature introduced here can be interpreted as the coherent component of the inverse of von Neumann-based temperature for the special case $C=C_r$. Indeed, we can conclude from Eq.~\eqref{cohr} that
\begin{equation}
    \beta=\beta_I-\beta_C,
\end{equation}
where
\begin{equation}
\beta=\left(\frac{\partial S(\rho )}{\partial U}\right)_{\{\lambda_i\}},
\end{equation}
denote the inverse of von Neumann-based temperature, with its incoherent and coherent parts given, respectively, by
\begin{equation}
\beta_I=\left(\frac{\partial S(\rho_I )}{\partial U}\right)_{\{\lambda_i\}}
\end{equation}
and
\begin{equation}
\beta_C=\left(\frac{\partial C_r(\rho )}{\partial U}\right)_{\{\lambda_i\}}.
\end{equation}

As an application, let us discuss the introduced coherent quantities in the context of magnon states. For simplicity, we consider the resonant case $\mathbf{k}=\mathbf{k}_0$, where both $C_r$ and $C_{\ln}$ are given by
\begin{equation}
C_{nNmk_0}=\sum_{q\in Q} p(q)\ln{|C^n_q|}.
\end{equation}
For large values of $n$, $N$, and $m$ (in the thermodynamic limit), coherence reveals its additive nature:
\begin{equation}
C(n,m,N)= ns\left(\frac{m}{N}\right),
\end{equation}
where
\begin{equation}
    s(x)=-x\ln x-(1-x)\ln(1-x)
\end{equation}
denotes the binary Shannon entropy. In the dilute-magnon regime, the Hamiltonian of $q$ magnons in a subsystem composed of $n$ sites is effectively described by
\begin{equation}\label{hamiltonian2}
H_{eff}=U(q,\epsilon_0)\ket{\psi_{nq\textbf{k}_0}}\bra{\psi_{nq\textbf{k}_0}}
\end{equation}
where
\begin{equation}
U(q,\epsilon_0)=q\epsilon_0
\end{equation}
denotes the corresponding eigenvalue and $\epsilon_0$ the magnon energy given by Eq.~\eqref{emag} with $k=k_0$. From Eqs.~\eqref{maxred}, \eqref{intenerg}, and \eqref{hamiltonian2}, we obtain the internal energy:
\begin{equation}
U(n,N,m,\epsilon_0)=\sum_{q\in Q} p(q)U(q,\epsilon_0)= nu(N,m,\epsilon_0)
\end{equation}
where
\begin{equation}
u(N,m,\epsilon_0)=\frac{m\epsilon_0}{N}
\end{equation}
is the energy density. Note that the internal energy is also an extensive quantity. Taking $\epsilon_0$ as a fixed parameter, the coherence density is expressed exclusively as a function of the energy density:
\begin{equation}
s(u)=-\left(\frac{u}{\epsilon_0}\right)\ln \left(\frac{u}{\epsilon_0}\right)-\left(1-\frac{u}{\epsilon_0}\right)\ln \left(1-\frac{u}{\epsilon_0}\right).
\end{equation}
In this scenario, the problem is formally equivalent to the classical ensemble of $N$ independent two-level systems, each possessing energy $0$ or $\epsilon_0$ (a two-level Boltzmann gas)~\cite{Salinas2001, Kittel1980}. Consequently, the inverse of the coherence temperature as a function of $u$ is given by
\begin{equation}\label{beta-u}
\beta_C(u)=\left(\frac{\partial s}{\partial u}\right)_{\epsilon_0}=\frac{1}{\epsilon_{0}}\ln \left(\frac{\epsilon_0}{u}-1\right).
\end{equation}
Note that the coherence temperature is an intensive quantity, due to its invariance with respect to the number of sites $n$.  The dependence of the energy density on coherence temperature is obtained by inverting Eq.~\eqref{beta-u}:

\begin{equation}
    u(\mathcal{\beta_C})=\frac{\epsilon_0}{e^{\epsilon_0\beta_C}+1}.
\end{equation}
Finally, the heat capacity can be determined by differentiating the energy density with respect to the coherence temperature:

\begin{equation}
\mathcal{C}(\beta_C)=\left(\frac{\partial u}{\partial \mathcal{T}}\right)_{\epsilon_0}=\frac{\epsilon_0^2\beta_C^2\,e^{-\epsilon_0\beta_C}}{\left(1+e^{-\epsilon_0\beta_C}\right)^2}.
\end{equation}
As expected in a two-level Boltzmann gas, $u$ increases with $\mathcal{T_C}$ as more particles occupy the higher energy level, asymptotically approaching a maximum value when $\mathcal{T_C}\rightarrow \infty$. Furthermore, a hallmark of a system with an upper energy bound is the occurrence of negative temperature, which arises when a population inversion takes place, that is, when the higher energy level becomes more populated than the lower one. Another important point is that $\mathcal{C}$ exhibits the well-known \textit{Schottky anomaly}: a peak at $\mathcal{T}_C\sim \epsilon_0$, where energy absorption is most efficient, which is typical behavior of systems with a limited number of energy levels.

%%%%%%%%%%%%%%%%%%%%%%%%%%%%%%%%%%%%%%%%%%%%%%%
%%%%%%%%%%%%%---Conclusions---%%%%%%%%%%%%%%%%%
%%%%%%%%%%%%%%%%%%%%%%%%%%%%%%%%%%%%%%%%%%%%%%%
\section{Conclusions}
\label{V}
We have investigated quantum coherence induced by non-interacting magnons in a ferromagnetic spin chain described by the isotropic Heisenberg model. By evaluating the density operator of an arbitrary subsystem, we showed that coherence reaches its maximum in the pure single-magnon state. This finding enabled the introduction of a coherence-based quantity formally analogous to Boltzmann entropy, as well as the definition of effective temperature and heat capacity. Our results suggest the development of an alternative and thermodynamically analogous description of quantum coherence in many-body systems. In future work, we intend to explore coherence-based thermodynamic quantities in a broader class of quantum systems, including those with particle interaction, finite-temperature regimes, and open-system dynamics subject to decoherence. 

%%%%%%%%%%%%%%%%%%%%%%%%%%%%%%%%%%%%%%%%%%%%%%%
%%%%%%%%%%%---Acknowledgments---%%%%%%%%%%%%%%%
%%%%%%%%%%%%%%%%%%%%%%%%%%%%%%%%%%%%%%%%%%%%%%%
{\section*{Acknowledgments}}

We acknowledge helpful discussion with B. Marques and E. Novais. This research was supported by the Brazilian National Institute for Science and Technology of Quantum Information (INCT-IQ) and by the S\~ao Paulo Research Foundation (FAPESP), Brazil (process number 2023/16378-4). 

\bibliographystyle{apsrev4-2}
\bibliography{references} 

\end{document}